\documentclass[times,trackchanges]{aastex631} % twocolumn

\submitjournal{ApJL}

\usepackage{savesym}

\restoresymbol{SIX}{tablenum}
\usepackage{booktabs}
\usepackage{multirow}
\usepackage{physics}
\usepackage{textcomp}
\usepackage{hyperref}
\usepackage[capitalize]{cleveref}
\usepackage{amssymb}
\usepackage{xcolor}
\usepackage{graphicx}  % for \rotatebox
\usepackage{longtable} % 
\usepackage{makecell} % 
\usepackage{tabularx}
\usepackage{rotating}

\makeatletter
\usepackage{etoolbox}
\patchcmd\H@refstepcounter{\protected@edef}{\protected@xdef}{}{}
\makeatother

\graphicspath{{./}{assets/}}

\begin{document}

% ---- Front matter
\title{Switchbacks near Boundaries of Small-scale Magnetic Flux Ropes in the Young Solar Wind from Parker Solar Probe Observations}

\shorttitle{Switchbacks near Boundaries of SMFRs from PSP observations}
\shortauthors{Choi et al.}

% ---- Authors
\author[0000-0003-2054-6011]{Kyung-Eun Choi}
\affiliation{Space Sciences Laboratory, University of California Berkeley: Berkeley, CA, USA}
\author[0000-0001-6427-1596]{Oleksiy V. Agapitov}
\affiliation{Space Sciences Laboratory, University of California Berkeley: Berkeley, CA, USA}
\affiliation{Astronomy and Space Physics Department, National Taras Shevchenko University of Kyiv, Kyiv, Ukraine}
\author[0000-0001-9994-7277]{Dae-Young Lee}
\affiliation{Department of Astronomy and Space Science, Chungbuk National University, Cheongju, Republic of Korea}
\author[0000-0002-2011-8140]{Forrest Mozer}
\affiliation{Space Sciences Laboratory, University of California Berkeley: Berkeley, CA, USA}
\author[0000-0002-9954-4707]{Jia Huang}
\affiliation{Space Sciences Laboratory, University of California Berkeley: Berkeley, CA, USA}
\author[0000-0001-6016-7548]{Lucas Colomban}
\affiliation{Space Sciences Laboratory, University of California Berkeley: Berkeley, CA, USA}
\author[0000-0003-1138-652X]{Jaye L. Verniero}
\affiliation{Heliophysics Laboratory, NASA Goddard Space Flight Center, 8800 Greenbelt Road, Greenbelt, MD 20771, USA}
\author[0000-0003-2409-3742]{Nour Raouafi}
\affiliation{Applied Physics Laboratory, Johns Hopkins University, Laurel, MD, USA}
% ........ ..........

\correspondingauthor{Kyung-Eun Choi}
\email{kechoi@berkeley.edu}

% ---- Abstract
\begin{abstract}
The \textit{Parker Solar Probe} (PSP) mission has revealed frequent occurrences of switchbacks (SBs) and small-scale magnetic flux ropes (SMFRs) as prominent structures within the solar wind. These mesoscale features are observed across all heliocentric distances, with heightened activity in the young solar wind, such as successive SMFRs, blobs, and SBs using PSP \textit{in situ} observations. One study, in particular, focuses on SMFRs observed during the intervals of PSP co-rotating with the Sun, which suggests a similar source of the observed solar wind. In this letter, we identified SBs at the boundaries of SMFRs as a regularly observed phenomenon and found instances where SBs and SMFRs co-occur, with the significance level $\alpha<0.05$. The SMFR-related SBs - observed at the leading and trailing edges of an SMFR - exhibit well-organized axial co-orientations, with their polarity flipping, meaning the radial direction remains constrained while the transversal field reverses.
Furthermore, the axial field directions of SMFRs-related SBs appear to be more closely connected than to another SB that is spatially closer and are linked to the SMFR orientation. Our analysis of their relative geometry, which examines the alignment between SBs and the SMFR axis, reveals a distinct tendency emphasizing their correlation, further supporting the idea that the axes of SMFR-related SBs are presumably determined by the SMFR orientation. Observations suggest that a fraction of SBs is spatially and temporally associated with SMFRs, implying that processes related to SMFR boundaries may contribute to SB formation, or that SBs tend to develop in magnetic environments shaped by SMFRs.   
\end{abstract}


% ---- Main text
\section{Introduction}\label{sec:intro}

The \textit{Parker Solar Probe} \citep{fox+2016, raouafi+2023} mission has revealed frequent occurrences of switchbacks \citep[SBs -- sharp magnetic field deflections accompanied by radial velocity spikes,][]{bale+2019, kasper+2019} as key structures within the solar wind. These structures have been reported by many authors \citep[e.g.,][]{kahler_lin_1994,neugebauer+1995, kahler+1996,crooker+2004, owens+2013, huang+2017, horbury+2018, woolley+2021}, particularly in the young solar wind \citep[e.g.,][]{bale+2019, kasper+2019, horbury+2020, dudok_de_wit+2020, mozer+2020, froment+2021}. 

However, the origin and generation mechanisms are still under debate \citep{raouafi+2023}, with two prevailing perspectives: a solar origin or local generation within interplanetary space. The solar origin hypothesis suggests that switchbacks form due to processes occurring near the Sun, including surface-driven mechanisms such as photospheric flows (i.e., granulation and supergranulation) and network magnetic field evolution \citep{bale+2021, bale+2023, kasper+2021,de_pablos+2022, fargette+2022, shi+2022}, small-scale jet-like eruptions like spicules or coronal jets injecting plasma into the solar wind \citep{sterling_and_Moore2020,neugebauer_Sterling_2021,telloni+2022,lee+2024, bizien+2025}, and interchange reconnection at the boundary between open and closed magnetic field regions \citep{yamauchi+2004,fisk_kasper_2020,horbury+2020,sterling_and_Moore2020,zank+2020,bale+2021,drake+2021,he+2021,wyper+2022,huang+2023,rivera+2024}. \cite{mozer+2021} suggested that SBs form in transition regions near the Sun, further supporting the idea that their origin may be linked to dynamic changes in solar wind properties close to the solar surface. Unlike the solar origin hypothesis, the \textit{in situ} formation of SBs gains credibility from the findings of \cite{jagarlamudi+2023}, who reported that their occurrence increases with heliocentric distance. 

Various mechanisms may also generate switchbacks within interplanetary space. Alfvén waves and firehose-like instabilities have been proposed as possible drivers of switchback formation \citep{larosa+2021}. Additionally, switchbacks could evolve above the Alfvén point due to dynamic instabilities in the expanding solar wind \citep{schwadron_and_McComas_2021}. Another mechanism involves the expansion of the solar wind coupled with turbulence-driven processes, which can stretch and amplify pre-existing magnetic field fluctuations, potentially leading to switchback formation \citep{landi+2006, matteini+2015, tenerani_velli_2018, tenerani+2020, ruffolo+2020, squire+2020, squire+2022, mallet+2021, schwadron_and_McComas_2021,martinovic+2021, shoda+2021,toth+2023}.

Distributions and dynamics of the rotation angle of SBs have been studied extensively \citep{horbury+2020,macneil+2020,fargette+2022,laker+2022,meng+2022}. SBs tend to rotate in a clockwise direction relative to the Parker spiral in the ecliptic plane in the quiet solar wind as reported based on the measurements from the first PSP encounters \citep{meng+2022,fargette+2022} or Helios \citep{macneil+2020}. \cite{horbury+2020} and \cite{laker+2022} showed that SBs predominantly deflect in the tangential ($T$) direction, rather than the normal ($N$) direction, in the $RTN$ coordinates. Here, $R$ indicates the radial direction from the Sun to PSP, $T$ is defined as the cross product of the solar rotation axis and the $R$ axis, and $N$ results from the third vector completing the right-handed orthogonal rule. Moreover, they observed that successive SBs often maintain similar orientations over several hours. \cite{laker+2023} reported a coherent deflection signature along arc polarized on the $T$-$N$ plane, suggesting an association with proton core enhancements. \cite{agapitov+2023} demonstrated a relationship between the deflection angle and the Alfvénicity of SBs, emphasizing the constraint of SB's Alfv\'enicity. 
%These findings collectively suggest that the deflection angle of SBs is closely linked to their underlying physical properties, such as Alfvénicity and large-scale magnetic field structure.

Another mesoscale structure in the solar wind is small-scale magnetic flux ropes \citep[SMFRs,][]{Moldwin+1995}, which are prominent structures within the heliosphere. Although SMFRs have been frequently observed \citep[e.g.,][]{yu+2016,hu+2018,chen+2019,choi+2021}, particularly in the inner heliosphere \citep[e.g.,][]{cartwright_moldwin_2010, zhao+2020, zhao+2021, chen+2021, chen+2023, chen_Hu_2022, drake+2021, shi+2021, eastwood+2021, farooki+2024}, their origins and structural properties continue to be explored. Recently, \cite{huang+2023_N} reported that the occurrence rate of small ejecta from the solar surface is comparable to the detection rate of SMFRs, supporting their solar origin \cite{rouillard+2009, rouillard+2011}. Alternatively, the helicity of in-situ SMFRs (i.e., chirality, such as left-handed or right-handed), as derived from modeling, has not exhibited a clear pattern \citep{choi+2022,chen_hu_2025}. This contrasts with large-scale flux ropes, whose helicity is consistently aligned between in-situ observations and solar filaments \citep[e.g.,][]{rust1994}, with a dependence on the solar cycle, supporting their solar origin. Furthermore, the lack of a systematic helicity preference in SMFRs suggests that they may not share the same solar origin and could instead originate from interplanetary processes \citep[e.g.,][]{moldwin2000, cartwright_moldwin_2008, greco+2009, tian+2010, feng+2015, telloni+2016, zheng_hu_2018, sanchez_diaz+2019, lavraud+2020, reville+2020}. This approximately even distribution of helicity may also indicate the presence of the counter-helicity type within SMFRs \citep{jia+2024}, adding complexity to their structural characteristics.

Several potential relationships between SBs and SMFRs have been proposed: (i) SMFRs may generate SBs \citep{sterling_and_Moore2020}, (ii) SBs may give rise to SMFRs \citep{shi+2024}, and (iii) SBs may be observed either within SMFRs \citep{drake+2021,chen+2021,agapitov+2022} or in the regions between them \citep{choi+2024}. \cite{sterling_and_Moore2020} suggested coronal jets as a source of SBs, arguing that erupting mini-filament flux ropes create Alfvénic fluctuations, which steepen into SBs during propagation. Recently, \cite{shi+2024} suggested, through a 3D MHD simulation, the possibility that the reconnection between two SBs could generate a flux rope. \cite{drake+2021} and \cite{agapitov+2022} demonstrated that flux ropes formed through interchange reconnection exhibit signatures consistent with SBs. \cite{chen+2021} investigated SMFRs that overlap with SBs and suggested that some switchbacks may be detected when the spacecraft traverses flux-rope-like structures, providing observational support for this theory.
A recent study by \cite{choi+2024} presented sequential observations of low-$\beta$ SMFRs, which are embedded by blobs and distributed radially during PSP's co-rotational orbits with the Sun, following a radial alignment within a narrow longitudinal zone. They reported that these flux rope structures exhibited relatively low fluctuations compared to the surrounding solar wind and the presence of SBs in between these flux ropes. 

While \cite{choi+2024} has predominantly centered on the properties and behaviors of SMFRs, this study shifts the focus to the relationship between SMFRs (gray area) and SBs (light blue area), focusing on SB properties and their distribution to better understand SBs' association with SMFRs and shared origins.
In this letter, we present the notable patterns of SBs occurring at the boundaries of low-$\beta$ SMFRs based on published SB databases by \cite{huang+2023} and \cite{agapitov+2023}. Our results reveal well-organized relative orientations of leading and tailing SBs' axes in these cases, with the SBs exhibiting directional flipping of the magnetic field, meaning the radial direction remains constrained while the transversal magnetic field reverses within the SBs. Especially, this tendency is pronounced at the leading and trailing edges of SMFRs, suggesting a systematic interaction between SBs and SMFR boundaries.

\section{Event Overview and Data Descriptions}\label{sec:overview}

\begin{figure*}
    \centering
    \includegraphics[width=\textwidth]{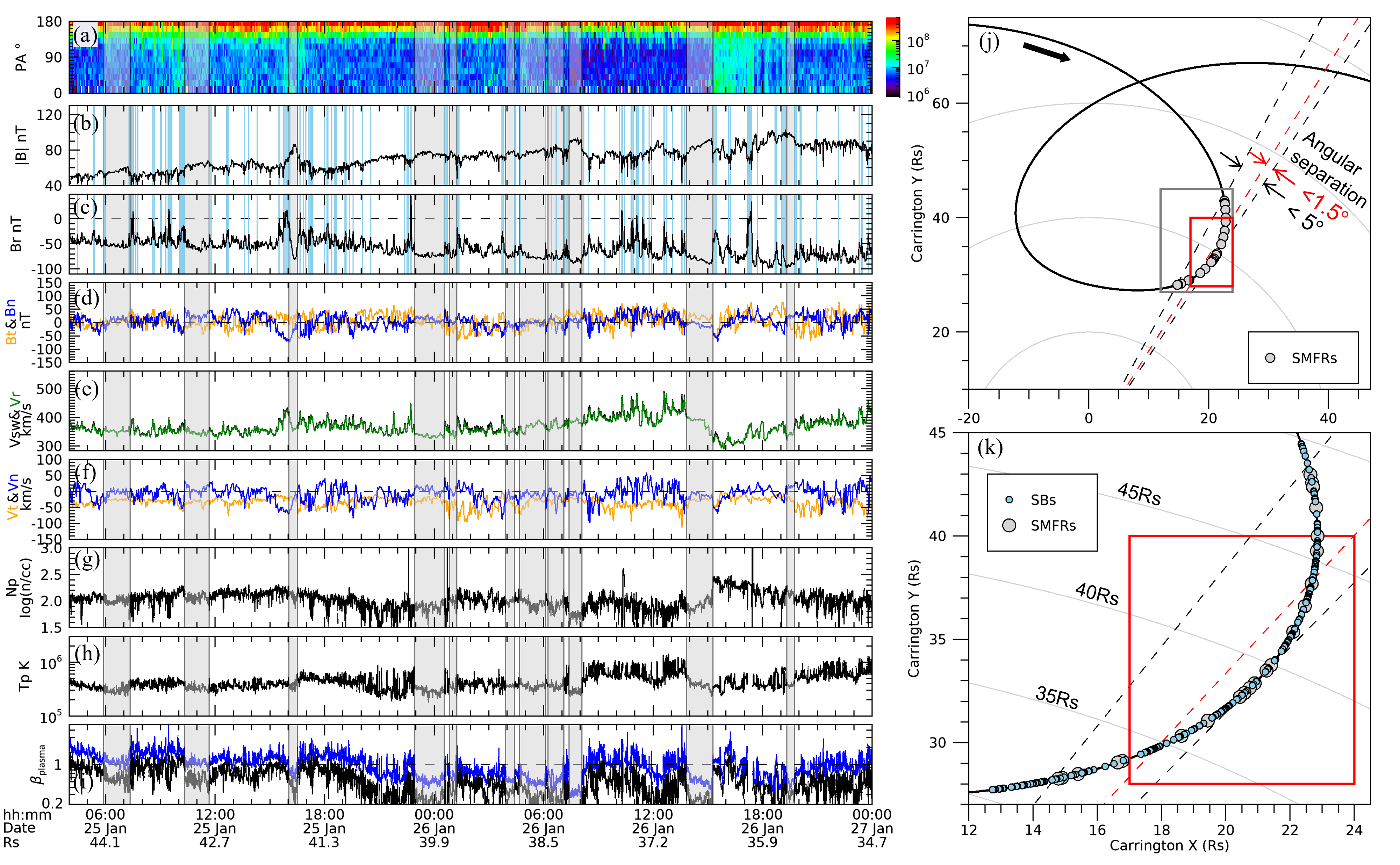}
    \caption{Overview of the SMFRs and SBs observed during the co-rotational intervals of PSP's E4 from 2020 January 25 to January 27. (a)-(i) From top to bottom, the panels show the pitch-angle distribution of suprathermal electron (314.5 eV), magnetic field intensity $|B|$ and $RTN$ components ($B_r, B_t, B_n$), solar wind speed ($V_{sw}$) and its $RTN$ components ($V_r, V_t, V_n$), proton density ($N_p$), proton temperature ($T_p$), and plasma $\beta$ ($\beta_{plasma}$) based on proton (black) and electron (blue). The gray shadings denote SMFRs, and the light blue shadings in (b) and (c) indicate SBs. (j) The PSP trajectory in Carrington coordinates with 11 SMFRs, as marked by gray-colored circles. (k) The locations of the 11 SMFRs and 245 SBs (22 SMFRs and 266 SBs), within $1.5^o$ ($5^o$) of longitudinal region ranged by dashed lines, corresponding to a red box (a gray box) in (j)}
   \label{fig:overview}
\end{figure*}

We focus here on PSP's Encounter 1 (E1) and Encounter 4 (E4), when PSP observed a series of SMFRs while the PSP’s azimuthal speed closely matched the solar rotation for 106 and 84 hours, respectively. Figures \ref{fig:overview}(b) and \ref{fig:overview}(c) illustrate the longitudinal variation and spatial distribution of SMFRs and SBs during the inbound co-rotating interval of E4. Across the total 190-hour intervals for E1 and E4, encompassing 36 low-$\beta$ SMFRs and 343 SBs, we infer a common origin for each co-rotating interval from the same source region with
 %. This inference is further supported by the small Carrington 
 longitudinal variation of less than $1.5^o$. These observations span a heliocentric distance range of 34.7 to 45.7 $R_S$ (Solar radii). 
 
Figure \ref{fig:overview} presents a series of low-$\beta$ SMFRs detected within the co-rotating interval of PSP, indicated on the orbital trajectory of PSP in Figures \ref{fig:overview}(j) and \ref{fig:overview}(k). The interval contains multiple SB structures located between SMFRs \citep{choi+2024}. In these intervals, SMFRs in the PSP data were identified by visually analyzing the magnetic field and plasma properties obtained from the corresponding measurements. The magnetic field intensity is enhanced by approximately 20-40$\%$ compared to the average over a two-hour window surrounding each event; and smooth rotation of magnetic field direction is observed. The proton density (Figure \ref{fig:overview}(g)) and/or proton temperature (Figure \ref{fig:overview}(h)) are typically lower than the surrounding solar wind, consistent with transient SMFR events (highlighted by gray regions in \Cref{fig:overview}(a--i)). Furthermore, the power spectrum of magnetic field perturbations exhibits a notable reduction within SMFRs. While this specific feature is not visually presented in this paper, it aligns with broader observational trends and supports the classification of SMFR events. These features collectively signify transient flux rope structures in the young solar wind \citep[see][for details on previous event selection]{choi+2024}. In determining  SMFRs, the consideration of specific physical quantities, along with the characteristic features of the $B$-spectrum, provides the most effective approach in the young solar wind for distinguishing them from torsional Alfvén waves \citep[e.g.,][]{marubashi+2010}, which can appear as pseudo-magnetic flux ropes. 

Alfvénic background solar wind conditions \citep{chen+2021} and wave activities \citep{shi+2021} make determining the boundaries of SMFRs in the young solar wind challenging. To address this, we have optimized the boundaries of SMFRs (start and end times) by minimizing the root-mean-squared error (Erms $<$ 0.35) in the force-free modeling for the cases of plasma beta (the ratio of plasma and magnetic pressure) less than 1 ($\beta<1$), and incorporating plasma parameters into the analysis, enabling more precise and reliable event identification compared to the prior analyses in \cite{choi+2024}. The plasma beta inside SMFRs is not necessarily low, as low beta is not a defining criterion for SMFRs in general. However, it serves as a useful indicator for distinguishing pseudo-SMFRs. We have allowed plasma beta to be lower than that of the background solar wind and noted in Table 1 cases where model fitting was not appropriate due to $\beta\ge1$ within the SMFR or fitting failure. The magnetic field data used in this study are from the PSP Fluxgate Magnetometer (MAG) - the part of the FIELDS suite \citep{bale+2016}, and plasma data are from the SWEAP suite \citep[Solar Wind Electrons Alphas and Protons;][]{kasper+2016}, which includes the Solar Probe Cup \citep[SPC;][]{case+2020} and Solar Probe Analyzers \citep[SPAN-i and SPAN-e;][]{livi+2022,whittlesey+2020}. We use the SPC data collected during E1, while the SPAN-i was used for the fitted proton the interval from E4. The electron density is derived from the analysis of plasma quasi-thermal noise (QTN) spectrum measured by the FIELDS Radio Frequency Spectrometer (Pulupa et al. 2017; Moncuquet et al. 2020). Additionally, we use SPAN-e to process electron pitch-angle distribution (PAD) data.

SBs, denoted by blue vertical regions in \Cref{fig:overview}(b--c), were identified using published catalogs \citep{kasper+2019,huang+2023,agapitov+2023}. Remarkably often, during these intervals, SBs were located near the edges of SMFRs. In the pitch angle distribution of suprathermal electrons, a beam consistently appears at 180 degrees—indicating the Toward polarity of the interplanetary magnetic field throughout the interval during SBs, the beam sometimes broadens, as seen in \Cref{fig:overview}(a). A more detailed investigation of the spatial and temporal relationships between SBs and SMFRs is provided in Figure 2 and summarized in Table 1.

\section{Orientations of Switchbacks}\label{sec:orientation}

\begin{figure*}
    \centering
    \includegraphics[width=\textwidth]{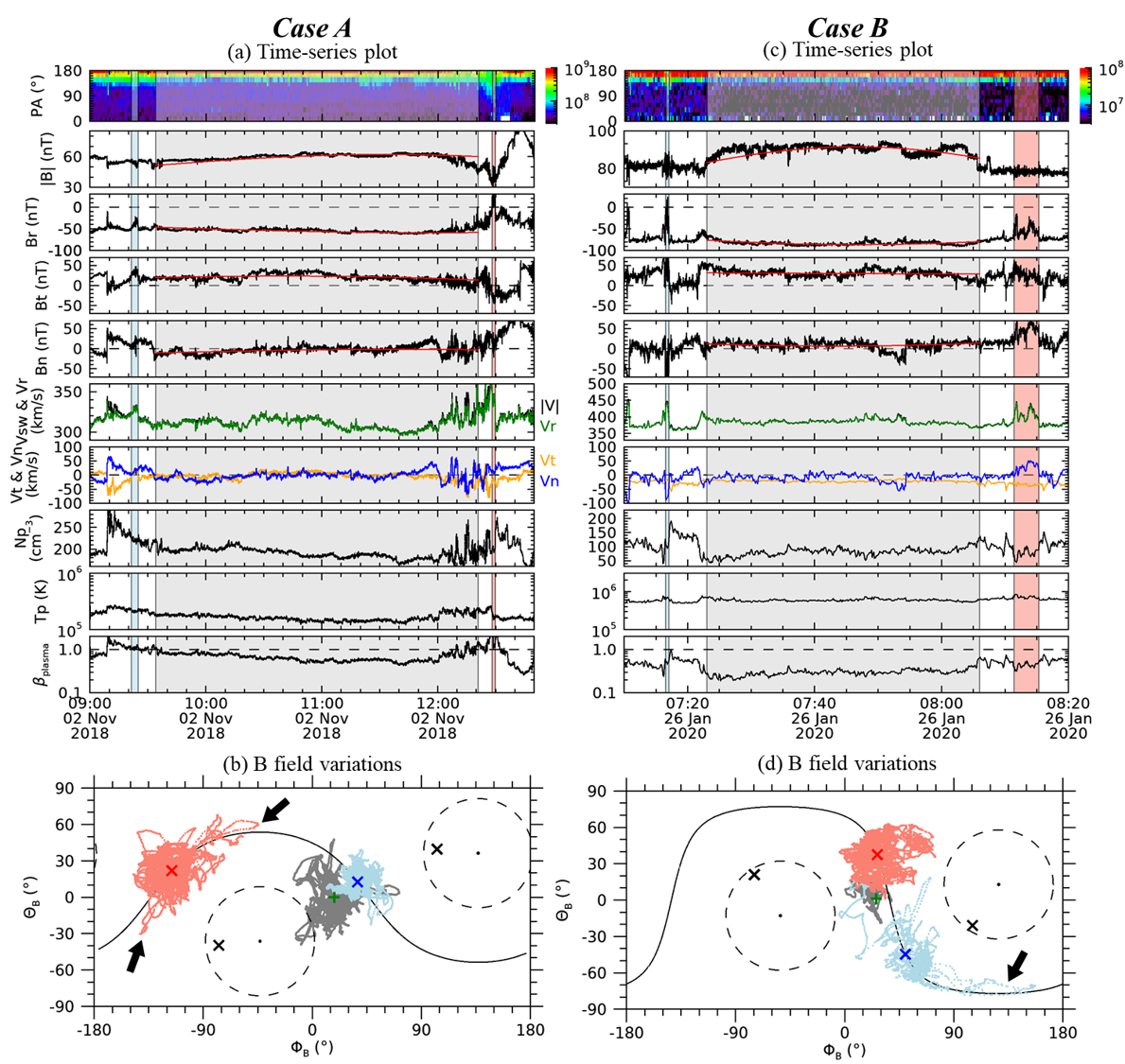}
    \caption{Examples of the SMFRs and SBs. Two cases represent $T$-flipping case (Case A, left column in (a) and (b)) and $N$-flipping case (Case B, right column in (c) and (d)). (a) and (c) show time series data from PSP, with panel order consistent with Figure 1. The red line represents the model fitting results. The shaded regions indicate different structures: light blue and light red shading mark the leading and trailing SBs, respectively, while the gray shading denotes the SMFR. (b) and (d) display the magnetic field data points in spherical coordinates, where the colors of the points follow the same scheme used for SBs and SMFRs in (a) and (c). The SMFR data in these panels are smoothed to emphasize the large-scale magnetic field rotation. $\Phi$ represents the azimuthal angle (longitude) of the magnetic field vector in the $RTN$ coordinate system, measured counterclockwise in the $R-T$ plane from the radial direction ($-R$), ensuring consistency with the sign of the tangential component ($T$). $\Theta$ represents the polar angle (latitude), measured from the $R-T$ plane. Cross symbols indicate specific directions:  blue and red denote the axes of the leading and trailing SBs (axes definition in the main text), and black corresponds to the SMFR axes from the force-free model. The black curve represents the plane passing through the two SBs, while the dashed curve marks the 45$^o$ reference. Green plus symbols represent the background solar wind. Black thick arrows indicate regions within the SBs that are aligned with the plane defined by the two SB axes.
    }
   \label{fig:exmaples}
\end{figure*}

\Cref{fig:exmaples} presents two SMFRs - Cases A and B - recorded during the E1 and E4, respectively, with SBs bounding each side of each SMFR. Within SMFRs, the radial component ($B_r$) is dominant, whereas it is significantly reduced or even reverses polarity during SBs. The tangential ($B_t$) and normal ($B_n$) components within SMFRs are consistent with the cylindrical model fitting (red curves in Figures \ref{fig:exmaples}(a) and \ref{fig:exmaples}(c)). 
The sign of leading and trailing SBs $B_t$ or/and $B_n$ changes when comparing the leading (blue) and trailing (red) SBs bounding each of SMFRs. Figures \ref{fig:exmaples}(b) and \ref{fig:exmaples}(d) further illustrate these spatial characteristics in spherical coordinates, showing distinct deflection of the magnetic field in the SBs. To clarify the large-scale rotation within the SMFRs, the magnetic field vectors in Figures \ref{fig:exmaples}(b) and \ref{fig:exmaples}(d) are smoothed within the SMFR intervals (gray lines). The fitted impact parameters, referring to the minimum distance between the spacecraft trajectory and the modeled flux rope axis, are 0.90 for Case A and 0.95 for Case B, indicating that both crossings were significantly off-axis (black crosses).

\subsection{\label{sec:level2}Case A: $T$-flipping Event on November 02, 2018}

The $T$-flipping event observed on November 2, 2018, at 45.1$R_S$, provides an example of how SBs alter the tangential ($T$) component of the magnetic field in the solar wind, during PSP's E1. This Case A event is characterized by a distinct reversal of the tangential component ($B_t$) within the leading and trailing SBs bounding the flux rope. Figures \ref{fig:exmaples}(a--b) illustrate the temporal and spatial characteristics of this event. The time series in Figure \ref{fig:exmaples}(a) presents the SMFR and its surrounding SBs, while the magnetic field map in Figure \ref{fig:exmaples}(b) shows the polarity changes in $B_t$.

In Figure \ref{fig:exmaples}(a), the SMFR, lasting approximately 3 hours, exhibits wave-like structures near its trailing edge ($\sim$1200 UT), accompanied by density fluctuations and a slight temperature enhancement. These features, located near the SMFR’s boundary, are consistent with similar observations reported by \cite{shi+2021}, where such wave-like structures were found near the edges of SMFRs. The leading and trailing SBs are positioned about 10 minutes apart from the SMFR's edges, a separation that may have arisen from the expansion of the SMFR or its surrounding structures during propagation. The leading SB is situated near the center of a blob-like structure with enhanced density, while the trailing SB is associated with a more perturbed region. Interestingly, the radial component ($B_R$) shows a clear reversal near the trailing SB, further emphasizing the localized polarity changes.

In Figure \ref{fig:exmaples}(b), the surrounding solar wind is dominated by the radial ($R$) direction, with a slight tangential ($T$) contribution (black cross symbol), consistent with Parker spiral characteristics. Following the definition of the magnetic axis in SBs, which treats them as magnetic tubes \citep{krasnoselskikh+2020}, we determined the SB axes (blue and red cross symbols) based on the average of the local magnetic field within each SB. A reference plane (black curve) is constructed geometrically from the leading and trailing SB axes and represents a geometric reference frame that captures their relative orientation. This plane is not fitted to the local magnetic field data but is defined solely by the two mean axis directions.
While the plane is defined in this geometric manner, the locally measured magnetic field vectors within each SB show a notable tendency to align with it. Although the magnetic field within each SB is expected to be distributed around its axis, the observed alignment with the plane defined by the two axes is noteworthy. Black thick arrows denote regions within the SBs where alignment with the geometrically defined plane is more pronounced.
This observation suggests that the plane reflects the local deflection of the magnetic field within each SB, indicating a strong geometric relationship between the two SBs despite the SMFR separating them. This provides a clue that the two SBs are dynamically or geometrically related. 

The SMFR axis was compared with the plane formed by the two SBs. Interestingly, the flux rope axis is tilted at an angle of approximately $63.2^o$ relative to this plane, indicating a significant inclination relative to the SB-defined plane. This notable tilt suggests a complex geometric relationship between the SB's plane and the SMFR's axes. While the exact nature of this relationship remains unclear, our preliminary analysis reveals intriguing trends that are further explored in Section 3.3.

Finally, a schematic of this configuration, shown in \Cref{fig:schematic}, will be discussed in detail in Section \ref{sec:dicussion_sum}, where potential mechanisms underlying these dynamics are also explored.

\subsection{\label{sec:level2}Case B: $N$-flipping Event on January 26, 2020}

The $N$-flipping event observed on January 26, 2020, during PSP's E4, provides another striking example of how switchbacks influence the solar wind's magnetic field. This Case B event occurred at a heliocentric distance of 38.3 $R_S$ and is characterized by an SMFR lasting approximately 43 minutes, with the leading and trailing SBs positioned about 10 minutes apart from the SMFR’s edges. The SMFR is embedded within blob-like structures, with the leading SB associated with a reversal in the radial component ($B_R$) and located just before a distinct blob structure. Interestingly, the two SBs exhibit similar density and temperature, indicating comparable plasma properties in their regions as shown in Figure \ref{fig:exmaples}(c).
In Figure \ref{fig:exmaples}(d), the data points within each SB closely follow the plane defined by the axes of the leading and trailing SBs, similar to the observations in Section 3.1. Surprisingly, the leading SB, represented by light blue points in Figure \ref{fig:exmaples}(d), closely follows the plane determined with the temporally more distant trailing SB, despite the presence of another closer SB around 07:10 UT. This suggests that the plane reflects the local deflection of the magnetic field within each SB, providing further evidence of a geometric relationship between the two SBs. The close alignment of solar wind properties with this plane implies that the two SBs could be magnetically connected as part of a single flux tube for this $N$-flipping case, potentially traversed twice by the spacecraft during its passage. This interpretation is supported by the schematic shown in \Cref{fig:schematic}.
When comparing the SMFR axis with the SB-defined plane, the flux rope axis is found to penetrate the plane at an angle of approximately 50.1$^o$, indicating a significant inclination. Although this relationship focuses on geometric differences compared to the $T$ flipping event discussed in Section 3.1, it also emphasizes the role of geometry in distinguishing the characteristics of $T$ and $N$ flipping events.
These findings imply that geometric factors could play a role in distinguishing $T$ and $N$-flipping events. 

%\subsection{\label{sec:level2}Basic Statistics for E1 and E4}
\subsection{\label{sec:level2}Basic Statistics for E4: Geometric properties depending on heliocentric distance}

%\begin{figure*}
%    \centering
%    \includegraphics[width=\textwidth]{assets/fig3_occr_success_SBs_new.png}
%    \caption{Occurrence rate and polarity distribution of SBs. (a)-(c) The occurrence rate of SBs with respect to the time difference ($dT_{SB-FR}$) from the nearest SMFR for E1 inbound (a), E1 outbound (b), and E4 inbound (c), respectively. Colored lines represent the occurrence rate obtained from the Monte Carlo experiments for within each interval, with significant levels of 50$\%$ (cyan), 68$\%$ (blue), and 97.5$\%$(magenta), respectively. (d)-(g) Polarity changes of adjacent SBs for E1+E4. The notations '1' and '2' on the axes denote the former and latter SBs, respectively. Distributions of changes in $T$ ((d) and (f)) and $N$ ((e) and (g)) show their relationship to SMFRs. The color bar in (d) and (e) represents the time difference between adjacent SBs ($dT_{SB-SB}$), while the legend in (f) and (g) indicates the flipping components. Filled symbols represent SMFR-related SBs (18 cases). The yellow shading in (f) and blue shading in (g) indicate $T$-flipping and $N$-flipping, respectively.}
%   \label{fig:fig3}
%\end{figure*}

\begin{figure*}
    \centering
    \includegraphics{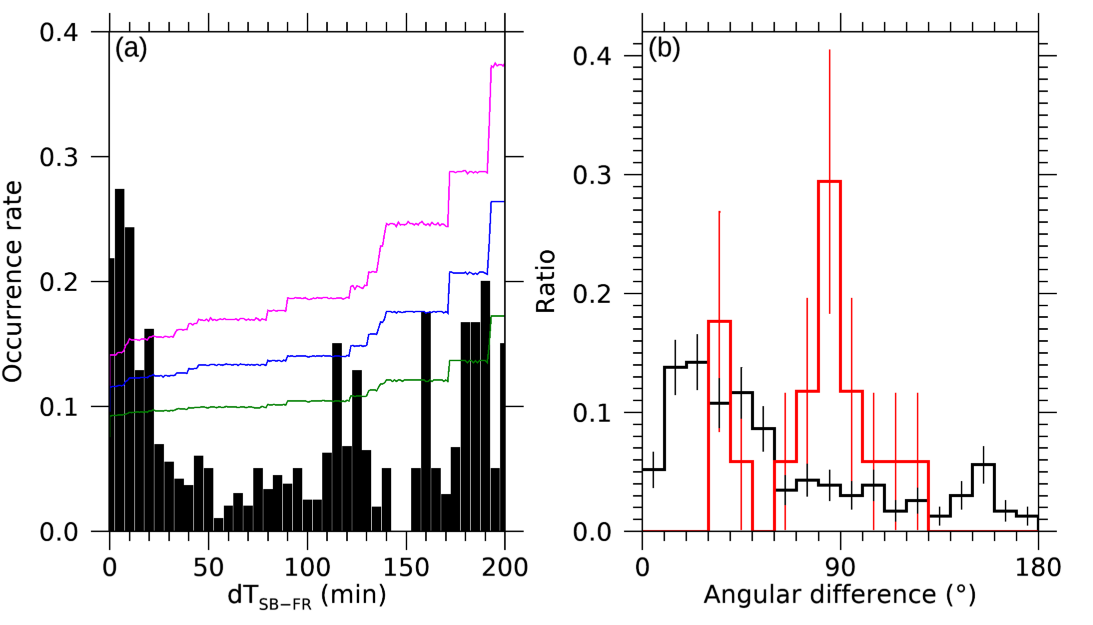}
    \caption{Occurrence rate and polarity distribution of SBs. (a) The occurrence rate of SBs with respect to the time difference ($dT_{SB-FR}$) from the nearest SMFR for E4 inbound. Colored lines represent the occurrence rate obtained from the Monte Carlo experiments within each interval, with significant levels of 50$\%$ (green), 68$\%$ (blue), and 97.5$\%$(magenta), respectively. (b) Histogram of angular differences between adjacent SBs in E4, categorized by FR association. The red bars represent SBs associated with SMFRs (17 cases), while the black bars indicate SBs not associated with SMFRs. Error bars represent the standard error (SE) of the mean for each group.
    }
   \label{fig:fig3}
\end{figure*}

In this section, we focus on E4 for the statistical analysis. This selection is guided by previous studies \citep{laker+2022}, which reported that certain PSP encounters, including E1 and E4, lack the asymmetry axis commonly observed in switchback deflections. This absence suggests a reduced influence of global magnetic structuring. In addition, E4 contains a particularly large number of well-defined SMFR events in the extended interval (see \Cref{fig:overview}(k)), offering a valuable dataset to investigate their geometric relationship with SBs under relatively neutral background conditions.

To ensure a consistent comparison, we begin by noting that during E1 and E4, a total of 343 SBs were identified within $1.5^o$ latitudinal zone. For the detailed analysis presented in this section, however, we concentrate on E4 extended range of $5^o$ in Carrington longitude depicted by the gray box in Figures \ref{fig:overview}(j) and \ref{fig:overview}(k) in this section. We analyze the 22 SMFRs and 266 SBs identified over the four-day interval within the range of 51.3 to 30.4 $R_S$. Figure \ref{fig:fig3}(a) provides the distribution of SBs as a function of their temporal proximity to SMFRs, revealing that the majority of SB-FR intervals ($dT_{SB-FR}$) are observed within 15 minutes. This suggests that SBs predominantly cluster around SMFR boundaries, implying a significant connection. Figure \ref{fig:fig3}(b) shows the deflection angles between successive SBs, differentiating cases where both leading and trailing SBs are present at SMFR boundaries from those without such associations. Specifically, we identify 17 SMFRs that are bounded by SBs on both sides, amounting to 34 SBs in total, depicted by red bars in Figure \ref{fig:fig3}(b), whereas those without SMFR-related boundaries are marked in black. The results indicate a significant difference in angular deflections, with SMFR-related SBs exhibiting larger angular changes than their non-SMFR counterparts. As noted in the introduction, previous studies \citep{horbury+2020,laker+2022} found that successive SBs tend to maintain similar orientations over several hours. Our result in Figure \ref{fig:fig3}(b) is consistent with those for non-SMFR-related SBs, showing the consistency in deflection angles between adjacent SBs. However, we found that SMFR-related SBs experience stronger deflections, potentially due to the local magnetic restructuring processes associated with SMFRs.

Table 1 provides a summary of the 17 SMFRs with SBs at both ends, accounting for 77$\%$ of the total 22 SMFRs identified in E4. This high fraction suggests that the presence of SBs at both ends of SMFRs is not a rare occurrence but rather a characteristic feature in this region. The table lists the start and end times of each SMFR, the success of the force-free model fitting, and detailed properties of the leading and trailing SBs, including their duration, time difference, angle difference, flipping patterns ($T$ and/or $N$), the angle $\theta_{FR-SB's plane}$, and distance from the Sun. Additionally, the angle $\theta_{FR-SB's plane}$ represents the inclination between the SMFR axis (determined by the force-free model fitting) and the plane containing the SBs at SMFR edges. 
By systematically comparing these cases as a function of heliocentric distance, we find that the $TN$ flipping pattern becomes more frequent at closer distances to the Sun. Furthermore, SMFRs exhibiting $TN$ flipping tend to have larger $\theta_{FR-SB's plane}$ values, indicating that the SBs in these cases form at a more inclined angle relative to the SMFR axis. E1 co-rotating interval ($<1.5^o$ latitudinal zone) has a similar trend provided by \Cref{tableA} in the appendix.  
%Among the 18 SMFRs, 15 SMFRs exhibit clear $T$ and/or $N$ flipping patterns, accounting for an impressive 83$\%$ of the cases. Notably, this represents 42$\%$ of all 36 SMFRs, significant of these patterns even within the whole SMFR population. The analysis further shows that all but one SMFR were successfully fitted with the force-free model, enabling accurate comparisons of their axes with the orientations of the bounding SBs. The parameter $\theta_{FR-SB’s plane}$ reveals distinct trends based on the observed flipping patterns. SMFRs exhibiting both $T$ and $N$ flipping patterns (3 cases) have an average angle of 63.3$^o$. In cases with only $T$ flipping (7 cases), the average angle is 49.1$^o$, while those with only $N$ flipping (5 cases) show an average of 51.5$^o$. By contrast, SMFRs without any flipping patterns (3 cases) exhibit a much smaller average angle of 15.2$^o$. These results demonstrate a clear correlation between the presence of $T$ and/or $N$ flipping patterns and larger $\theta_{FR-SB’s plane}$ angles.
%As a result, SMFRs associated with flipping patterns tend to have a stronger geometric or dynamic alignment with the SB's plane. Conversely, the smaller angles observed in non-flipping cases indicate a weaker or less structured alignment. 

% Please add the following required packages to your document preamble:
% \usepackage{booktabs}
% \usepackage{multirow}
\clearpage
\begin{figure*}[ht!]
    \centering
    \includegraphics[width=0.95\textwidth]{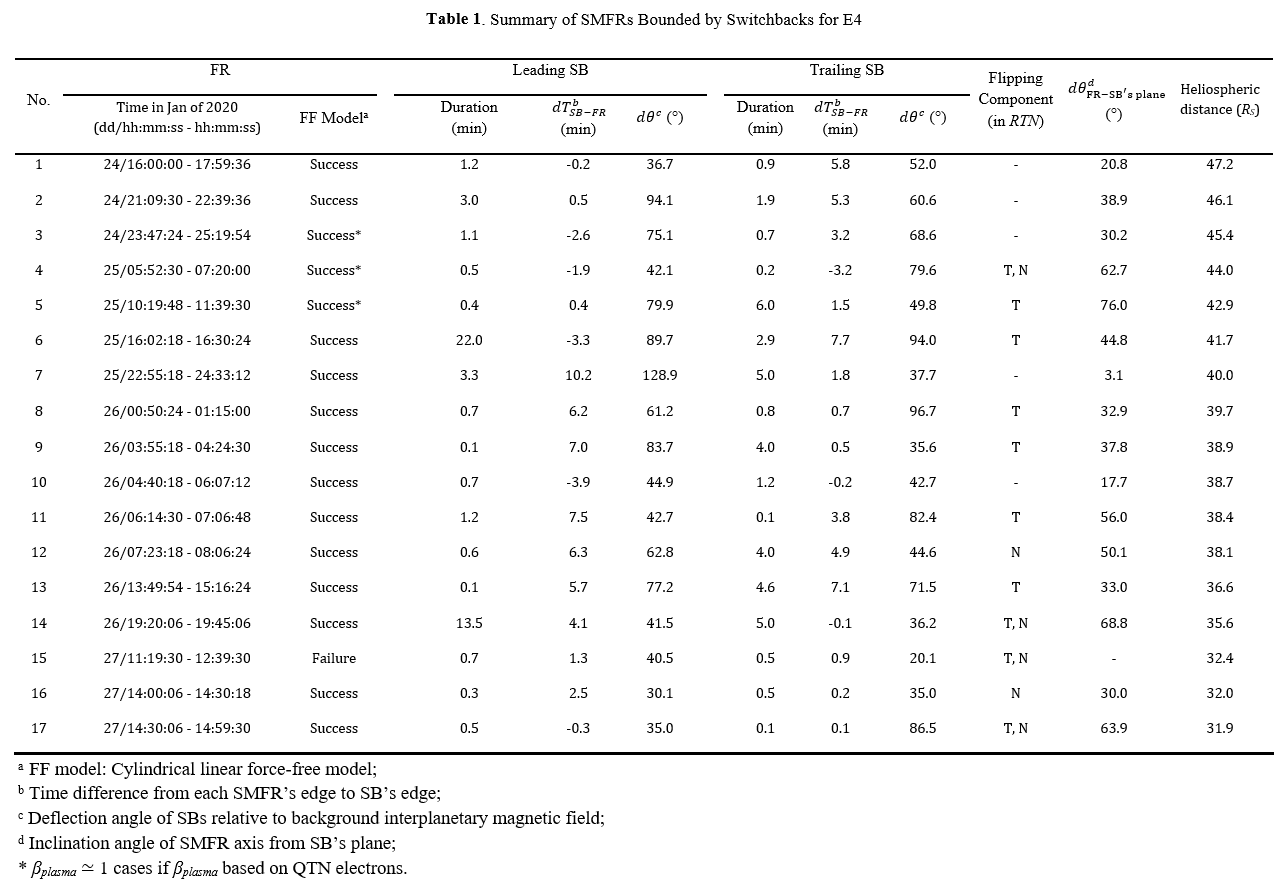}
\end{figure*}

\clearpage

\section{Discussion and Summary}\label{sec:dicussion_sum}

\begin{figure*}
    \centering
    \includegraphics[width=\textwidth]{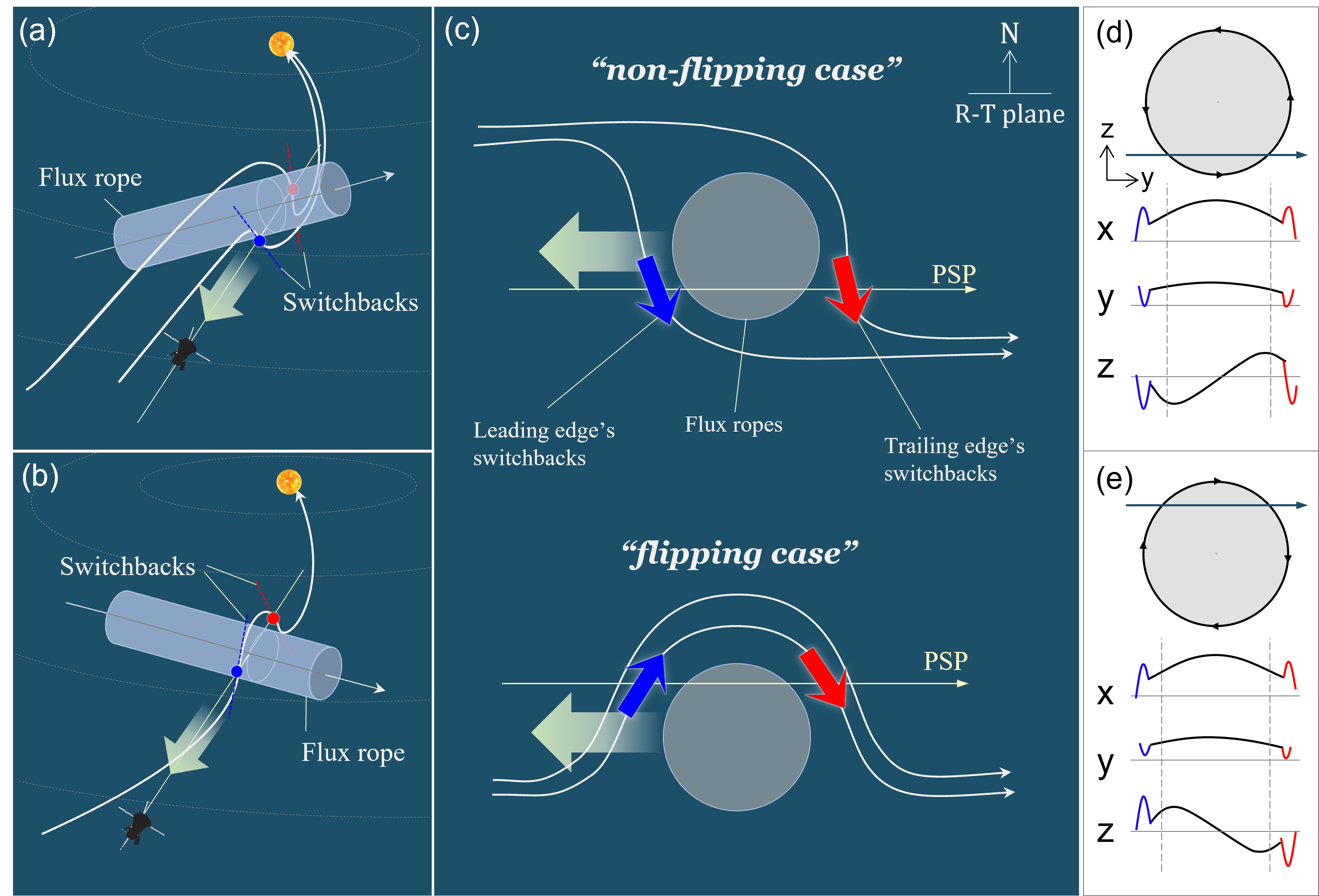}
    \caption{Schematic of SMFR-related SBs' magnetic field geometry and the corresponding SMFR. Panels (a) and (b) illustrate possible scenarios in which PSP traverses an SB-SMFR-SB structure over time. White curves represent magnetic field lines, and the cylinder represents the SMFR. Blue and red dots mark the detection points where the spacecraft encounters SBs. Panel (c) provides a cross-section view, where blue and red arrows indicate SBs and their respective directions. Panels (d) and (e) show examples of observed magnetic field components. Here, $x$ represents the SMFR axis direction, $y$ corresponds to the spacecraft's trajectory, and $z$ is the cross-product direction. For the two cases presented in Figure \ref{fig:exmaples}, the corresponding $RTN$ directions in these specific events are $x=T$, $y=-R$, and $z=N$, respectively. The blue, black, and red curves represent the leading SB, SMFR, and trailing SB, respectively.
    The schematic illustrates the spatial arrangement of SBs and SMFRs, not the temporal evolution of magnetic field lines.}
    %(a) SBs' axial magnetic field $T$-flipping case. The White curves represent magnetic field lines, and the cylinder represents the SMFR. The blue and red dots indicate the detection of SBs by the spacecraft as it passes through. (b) Same as (a), but for $N-$flipping case. 
    
   \label{fig:schematic}
\end{figure*}

In this letter, we present the results of the analysis of the relationship between small magnetic flux ropes (SMFRs) and switchbacks (SBs) in the young solar wind based on Parker Solar Probe (PSP) observations during Encounters 1 and 4. According to the statistical results for the SB deflection distributions in \cite{laker+2022}, SBs demonstrate a preference for the tangential direction of their axial magnetic field \citep[as illustrated in Figure 4 in][]{laker+2022}. While \cite{laker+2022} identified systematic magnetic configurations and specific symmetry axes in most of the PSP encounters (from the first 8 PSP encounters), they noted that Encounters 1 and 4 deviated from this trend, suggesting a possible lack of alignment with large-scale solar wind structures. Our analysis focuses on these 'exceptional' encounters, revealing that the more chaotic SBs orientation seen in Encounters 1 and 4 are, in fact, was connected to smaller-scale solar wind structures. We identified distinct connections (alignment) of SBs with the boundaries of SMFRs and reversal of leading and trailing SBs axial magnetic filed polarity patterns by examining their spatial and magnetic configurations. \Cref{fig:schematic} schematically illustrates possible cases of SB polarity flipping. In this figure, two SBs—referred to as the leading and trailing SBs—are shown on either side of an SMFR, represented by blue and red arrows, respectively.

The observation of SB deflection angles demonstrates the relationship between successive SBs, suggesting coherent magnetic behavior across multiple events. As emphasized in \cite{horbury+2020} and \cite{laker+2022}, successive switchbacks often maintain a consistent deflection angle over extended intervals, suggesting these coherent magnetic structures are shaped by large-scale solar wind dynamics. In contrast, the more recent findings of \cite{laker+2023} reveal cases of polarity reversals ($TN$ flipping) in successive SBs, indicating localized and potentially transient magnetic interactions. One of our results shows that during Encounter 4, $T$ and/or $N$ flipping patterns are observed in 12 cases (about 70.6$\%$ of the total 17 SMFRs), illustrating the significant occurrence of polarity reversals. Our study aligns more closely with the findings of \cite{laker+2023}, as we observe distinct $TN$ polarity flipping patterns. However, unlike \cite{laker+2023}, which reported these patterns in successive SBs over extended regions, we identify such behaviors specifically near SMFR boundaries. 

Given the narrow longitudinal detection region, it is clear that these features originate from the same coronal structures, which can be traced back to the source surface. However, subtle variations in these structures may account for the observed differences in their properties and spatial configurations. It is also worth emphasizing that the SMFRs examined in this study are characterized by their transient nature and low-$\beta$. Furthermore, while some SMFR-related SBs occur within the same series of SMFRs, their properties are not uniform. The distance between SBs and SMFR boundaries, the obliqueness of SB orientations relative to the SMFR axes, and their characteristics also vary with heliospheric distance, as mentioned in \Cref{sec:level2}. In general, SMFRs and SBs are influenced by the background solar wind conditions and the surrounding magnetic environment, with their properties varying depending on factors such as heliocentric distance and solar cycle phase. However, due to the limitations of the detection region, finer structural details may remain unresolved. To better understand these variations, it is necessary to examine a broader range of heliocentric distances and different phases of the solar cycle. Further statistical analysis and simulations incorporating diverse solar wind conditions will be essential for a more comprehensive interpretation.

Extending the results reported in \cite{laker+2022, laker+2023}, our study highlights how SMFR boundary affects the behavior of SB deflection angles. Furthermore, the suggested cogeneration of SBs and SMFRs, which could emerge from interactions between multiple switchbacks, as shown in recent three-dimensional MHD simulations \citep{shi+2024} - suggests a tightly coupled relationship between these structures. In summary, our key findings in this letter are:

\textbf{(1)} Most of 22 SMFRs recorded during PSP Encounter 4 were surrounded by SBs near the SMFR boundary: SBs were present at 86.4$\%$ of the SMFR boundaries and 77.3$\%$ of the SMFRs were bounded by SBs at the leading and trailing edges. 
%\ke{4 SMFRs out of 5 SFMRs have SBs on one side. 34 SBs + 4 SBs / 44 boundaries = 86.4$\%$}

\textbf{(2)} 35.9$\%$ of the 266 SBs observed during PSP Encounter 4, demonstrated statistically significant (with the significance level $\alpha<0.05$) relation to the SMFR boundary - in the 15-minute vicinity. In contrast, 64.1$\%$ of the SBs were randomly distributed in the inter-SMFR regions. 
%\ke{57.0$\%$ is outside; 7.1$\%$ is inside of SMFRs}

\textbf{(3)} Successive SBs associated with SMFRs showed systematic variations in their axial magnetic field direction. In particular, a large majority (71$\%$) of SMFR-related SBs exhibited $T$ and/or $N$ component polarity reversal patterns between the leading and trailing SBs.

\section{Acknowledgments}
KEC and OVA were supported by NASA contracts 80NSSC22K0522, 80NSSC22K0433, 80NNSC19K0848, 80NSSC21K1770, and NASA’s Living with a Star (LWS) program (contract 80NSSC20K0218). 
We thank the NASA Parker Solar Probe Mission, the SWEAP team led by Justin Kasper, and the FIELDS team led by Stuart Bale for the use of the data.

\clearpage
\appendix

\section{Geometric properties of SMFR-related SBs for E1}

Table A summarizes 7 SMFRs from the inbound and outbound corotating interval of E1, out of a total 25 SMFRs, and shows that the occurrence of $TN$ flipping is associated with larger $\theta_{FR-SB's plane}$ values, consistent with the trend observed in E4. 

% Please add the following required packages to your document preamble:
% \usepackage{booktabs}
% \usepackage{multirow}

\begin{figure*}[ht!]
    \centering
    \includegraphics[width=0.95\textwidth]{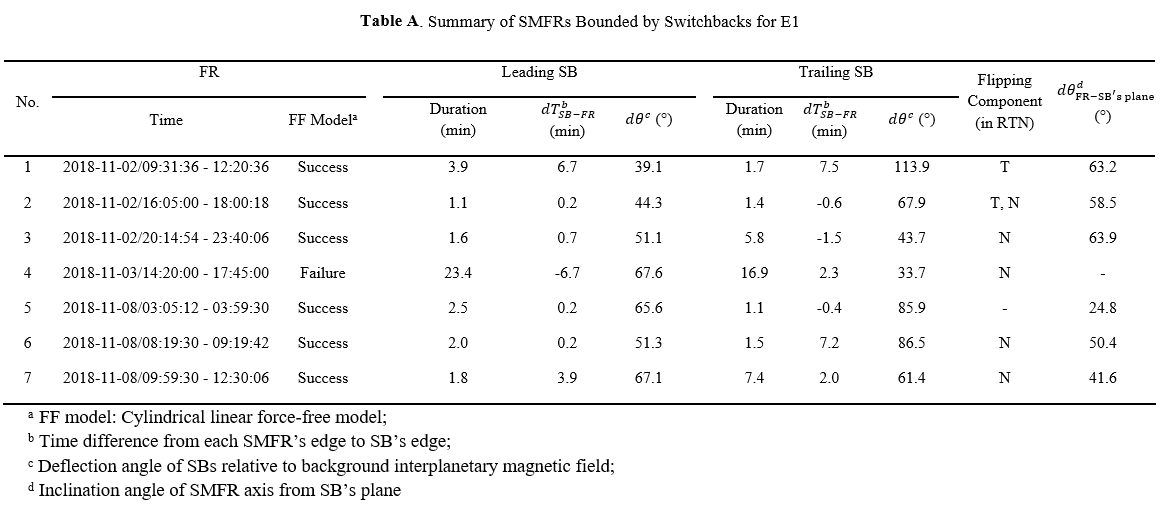}
\end{figure*}

\clearpage

\bibliographystyle{aasjournal}
\bibliography{ref}

\end{document}